\documentclass[sigconf,nonacm,natbib=false,screen]{acmart}

\usepackage[utf8]{inputenc}
\usepackage[T1]{fontenc}
\usepackage[american]{babel}

\usepackage[frozencache]{minted}
\usepackage{csquotes} %

\usepackage{booktabs}
\usepackage{listings}
\usepackage{enumitem}
\usepackage{amsmath}
\usepackage{subcaption}
\usepackage{tabularx}
\usepackage{makecell}
\usepackage{environ}

\usepackage[binary-units]{siunitx}
\DeclareSIUnit{\nothing}{\relax}
\DeclareSIUnit{\x}{\times}
\useshorthands*{"}
\defineshorthand{"-}{\babelhyphen{hard}}

\PassOptionsToPackage{
	sortcites,
	maxcitenames=2,
	maxbibnames=8,
}{biblatex}
\usepackage{biblatex}
\bibliography{main}

\usepackage{xcolor}
\definecolor{ETHa}{RGB}{31,64,122}      %
\definecolor{ETHb}{RGB}{72,90,44}       %
\definecolor{ETHc}{RGB}{18,105,176}     %
\definecolor{ETHd}{RGB}{114,121,28}     %
\definecolor{ETHe}{RGB}{145,5,106}      %
\definecolor{ETHf}{RGB}{111,111,100}    %
\definecolor{ETHg}{RGB}{168,50,45}      %
\definecolor{ETHh}{RGB}{0,122,150}      %
\definecolor{ETHi}{RGB}{149,96,19}      %

\PassOptionsToPackage{pdfpagelabels=false}{hyperref}
\usepackage{hyperref}

\hypersetup{%
	colorlinks=true,%
	urlcolor=ETHg,%
	linkcolor=ETHc,%
	citecolor=ETHd,%
	breaklinks=true,%
	pageanchor=true,%
}

\makeatletter
\DeclareRobustCommand{\varname}[1]{\begingroup\newmcodes@\mathit{#1}\endgroup}
\makeatother

\begin{document}

\title{Rumble: Data Independence for Large Messy Data Sets}
\subtitle{~}

\author{Ingo Müller}
\orcid{0000-0001-8818-8324}
\affiliation{%
  \institution{ETH Zurich}
  \streetaddress{Stampenbachstrasse 114}
  \city{Zurich}
  \state{Switzerland}
  \postcode{8057}
}
\email{ingo.mueller@inf.ethz.ch}

\author{Ghislain Fourny}
\orcid{0000-0001-8740-8866}
\affiliation{%
  \institution{ETH Zurich}
  \streetaddress{Stampenbachstrasse 114}
  \city{Zurich}
  \state{Switzerland}
  \postcode{8057}
}
\email{ghislain.fourny@inf.ethz.ch}

\author{Stefan Irimescu}
\authornote{The contributions of these authors were made
            during their studies at ETH Zurich.}
\affiliation{%
  \institution{Beekeeper AG}
  \city{Zurich}
  \state{Switzerland}
}
\email{stefan.irimescu@beekeeper.io}

\author{Can Berker Cikis}
\authornotemark[1]
\affiliation{%
  \institution{(unaffiliated)}
  \city{Zurich}
  \state{Switzerland}
  \postcode{8057}
}
\email{canberkerwork@gmail.com}

\author{Gustavo Alonso}
\affiliation{%
  \institution{ETH Zurich}
  \streetaddress{Stampenbachstrasse 114}
  \city{Zurich}
  \state{Switzerland}
  \postcode{8057}
}
\email{alonso@inf.ethz.ch}
\email{}
\email{}

\begin{abstract}
This paper introduces Rumble, a query execution engine
for large, heterogeneous, and nested collections of JSON objects
built on top of Apache Spark.
While data sets of this type are more and more wide-spread,
most existing tools are built around a tabular data model,
creating an impedance mismatch for both the engine and the query interface.
In contrast, Rumble uses JSONiq,
a standardized language specifically designed for querying JSON documents.
The key challenge in the design and implementation of Rumble
is mapping the recursive structure of JSON documents and JSONiq queries
onto Spark's execution primitives based on tabular data frames.
Our solution is to translate a JSONiq expression into a tree of iterators
that dynamically switch between local and distributed execution modes
depending on the nesting level.
By overcoming the impedance mismatch \emph{in the engine},
Rumble frees the user from solving the same problem for every single query,
thus increasing their productivity considerably.
As we show in extensive experiments,
Rumble is able to scale to large and complex data sets in the terabyte range
with a similar or better performance than other engines.
The results also illustrate that Codd's concept of data independence
makes as much sense for heterogeneous, nested data sets
as it does on highly structured tables.
\end{abstract}

\keywords{JSONiq, Spark, RDD, DataFrames, distributed parallel processing,
  SQL, document stores, JSON, nested data, heterogeneous data, NoSQL}

\maketitle

\pagestyle{plain}
\begingroup
\renewcommand\thefootnote{}\footnote{\noindent
  This work is licensed under the Creative Commons
  Attribution-NonCommercial-NoDerivatives (BY-NC-ND) 4.0
  International License.
  To view a copy of this license,
  visit \url{http://creativecommons.org/licenses/by-nc-nd/4.0/}.
  For any use beyond those covered by this license,
  obtain permission by contacting the authors.
  Copyright is held by the owner/author(s).
}\addtocounter{footnote}{-1}
\endgroup

\vspace{2cm}
\section{Introduction}

JSON is a wide-spread format for large data sets.
Its popularity can be explained by its concise syntax,
its ability to be read and written easily by both humans and machines,
and its simple, yet flexible data model.
Thanks to its wide support by programming languages and tools,
JSON is often used to share data sets between users or systems.
Furthermore, data is often first produced in JSON,
for example, as messages between system components or as trace events,
where its simplicity and flexibility
require a very low initial development effort.

However, the simplicity of creating JSON data on the fly
often leads to a certain heterogeneity of the produced data sets,
which creates problems on its own.
As a running example, consider the GitHub Archive,
a public data set of about \SI{1.1}{\tera\byte} of compressed JSON objects
representing about 2.9 billion events
recorded by the GitHub API between early 2011 and today.
The data set combines events
of different types and different versions of the API---%
presumably, because each event is simply archived as-is.
Each individual event in the data set is already complex,
consisting of nested arrays and objects
with typically several dozen attributes in total.
However, the variety of events
makes the collection as a whole even more complex:
all events \emph{combined}
have more than \SI{1.3}{\kilo\nothing} different attributes.
Furthermore, about \SI{10}{\percent} of these attributes have mixed JSON types.
Analyzing this kind of data sets thus requires dealing with ``messy'' data,
i.e., with absent values, nested objects and arrays, and heterogeneous types.

We argue that today's data analytic systems
have unsatisfactory support for large, messy data sets.
While many systems allow reading JSON-based data sets,
they usually represent them in some flavor of data frames,
which often do not work with nested arrays (for example, \texttt{pandas})
and almost always only work with homogeneous collections.
Attributes with a mix of types are either dropped or kept as strings---%
including all nested attributes---%
and must be parsed manually before further analysis.
For example, because there is a tiny number of integers among the objects
in the \mintinline{xquery}{.payload.issue} path,
all values at this path are stored as strings.
Overall, this is the case for about a quarter of the data set,
which is hence inaccessible for immediate analysis.
Additionally, data-frame-based systems usually map
non-existing values \emph{and} \mintinline{xquery}{null} values%
\footnote{These are different concepts in JSON:
          The attribute \mintinline{xquery}{foo} is non-existent
          in \mintinline{JSON}{{"bar": 42}},
          but has the value \mintinline{JSON}{null}
          in \mintinline{JSON}{{"foo": null}}.}
to their own representation of \mintinline{SQL}{NULL},
making it impossible to distinguish the two
in the (admittedly rare) cases where this might be necessary.
In short, the conversion to a data frame
seems to be an inadequate first step for analyzing messy JSON data sets.

Furthermore, SQL is arguably not the most intuitive language
when it comes to nested data,
and the same is true for other languages
originally designed for flat data frames or relations.
While the SQL standard does define an array data type
and even sub-queries on individual arrays,
this is done with pairs of \mintinline{SQL}{ARRAY(.)}
and \mintinline{SQL}{UNNEST(.)},
which is more verbose and less readable then necessary.
For example,%
\footnote{Alternative syntax include \mintinline{SQL}{LATERAL JOIN},
          for which the same arguments applies.}
the following query
selects the elements \mintinline{SQL}{x}
within an array attribute \mintinline{SQL}{arr}
that are larger than 5 and multiplies them by 2:
\begin{minted}{SQL}
SELECT arr, ARRAY(
    SELECT x*2 FROM UNNEST(arr) AS x WHERE x < 5)
FROM table_with_nested_attributes;
\end{minted}
What is worse is that few systems
implement the array functionality of the standard even partially,
let alone all of it.
Instead, if they support arrays at all,
they do so through (often non-standard) functions
such as \mintinline{SQL}{array_contains}, which are less expressive.
For example, SQL can express ``argmin-style'' queries only with a self-join
and none of the SQL dialects we are aware of
has an equivalent of \mintinline{SQL}{GROUP BY} for arrays.
Consequently, queries on messy data sets
might consist of three different sub-languages:
(1) SQL for the data frame holding the data set,
(2) \mintinline{SQL}{array_*} functions for nested data,
and (3) user-defined functions for otherwise unsupported tasks
including parsing and cleansing the heterogeneous attributes.
Current tools are thus not only based on an inadequate data model
but also on a language with sub-optimal support for nested data.

In this paper, we present the design and implementation of \emph{Rumble},
a system for analyzing large, messy data sets.
It is based on the query language JSONiq~\cite{fourny2013,jsoniqBook},
which was designed to query collections of JSON documents,
is largely based on W3C standards, and has several mature implementations.
JSONiq's data model accurately represents any well-formed JSON document
(in fact, any JSON document \emph{is} automatically a valid JSONiq expression)
and the language uses the same, simple and declarative constructs
for all nesting levels.
This makes it natural to deal with all aspects of messy data sets:
absent data, nested values, and heterogeneous types.
Furthermore, the fact that JSONiq is declarative
opens the door for a wide range of query optimization techniques.
Thanks to JSONiq, Rumble does hence not suffer
from the short-comings of data-frame-based systems mentioned above
but instead promises higher productivity for users
working with JSON data.

The key difference of Rumble
compared to previous implementations of JSONiq
such as Zorba~\cite{zorba} and IBM WebSphere~\cite{WebSphere2017}
is the scale of data sets it targets.
To that aim, we implement Rumble on top of Apache Spark~\cite{sparkOriginal},
such that it inherits the big data capabilities of that system,
including fault tolerance,
integration with several cluster managers and file systems
(including HDFS, S3, Azure Blob Storage, etc)
and even fully managed environments in the cloud.
As a side effect, Rumble also inherits Spark's support
for a large number of data formats
(including Parquet, CSV, Avro, and even libSVM and ROOT,
whose data models are subsets of that of JSON).
In contrast, Zorba and WebSphere are designed for processing small documents
and only execute queries on a single core.
Rumble thus combines a high-level language
that has native support for the full JSON data model
with virtually unlimited scale-out capabilities,
enabling users to analyze messy data sets of any scale.
This also shows that the proven design principle of data independence
makes just as much sense for semi-structured data
as it makes for the relational domain.

\section{Background}
\label{sec:background}

\begin{figure}
\begin{minted}{JSON}
{
  "type": "PushEvent",
  "commits": [{"author": "john", "sha": "e230e81"},
              {"author": "tom", "sha": "6d4f151"}],
  "repository":
      {"name": "hello-world", "fork": false},
  "created_at": "2013-08-19"
}
\end{minted}
\caption{A (simplified) example JSON object from the Github Archive data set.}
\label{fig:json-example}
\end{figure}

\noindent\textbf{JSON.}~~%
JSON~\cite{json} (JavaScript Object Notation) is a syntax
describing possibly nested values.
Figure~\ref{fig:json-example} shows an example.
A JSON value is either a number, a string,
one of the literals \mintinline{xquery}{true},
\mintinline{xquery}{false}, or \mintinline{xquery}{null},
an array of values, or an object,
i.e., a mapping of (unique) strings to values.
The syntax is extremely concise and simple,
but the nestedness of the data model makes it extremely powerful.

Related are the JSON Lines~\cite{json-lines} and ndjson~\cite{ndjson} formats,
which essentially specify a collection of JSON values
to be represented with one value per line.

\smallskip\noindent
\textbf{JSONiq.}~~%
\label{sec:background:jsoniq}
JSONiq~\cite{jsoniq,fourny2013} is a
declarative, functional, and Turing-complete query language
that is largely based on W3C standards.
Since it was specifically designed for querying large quantities of JSON data,
many language constructs make dealing with this kind of data easy.

All JSONiq expressions manipulate
ordered and potentially heterogeneous ``sequences'' of ``items''
(which are instances of the JSONiq Data Model, JDM).
\emph{Items} can be either
(i) atomic values including all atomic JSON types
as well as dates, binaries, etc.,
(ii) structured items, i.e., objects and arrays,
or (iii) function items.
\emph{Sequences} are always flat, unnested automatically, and may be empty.

Many expressions work on sequences of arbitrary length,
in particular, the expressions navigating documents:
For example, in \mintinline{xquery}{$payload.commits[].author}, %
the object look-up operator \texttt{.}
is applied to any item in \mintinline{xquery}{$payload} %
to extract their \mintinline{xquery}{commits} member
and the \texttt{[]} operator is applied to any item in the result thereof
to unnest their array elements as a sequence of items.
Both operators return an empty sequence
if the left-hand side is not an object or array, respectively.
The result, all array elements of all \mintinline{xquery}{commits} members,
is a single flat sequence, which is consumed by the next expression.
This makes accessing nested data extremely concise.
Some expressions require a sequence to be of at most or exactly one element
such as the usual expressions
for arithmetic, comparison, two-valued logic, string manipulation, etc.

Where it makes sense, expressions can deal with absent data:
For example, the object look-up simply returns
only the members of those objects that do have that member,
which is the empty sequence if none of them has it,
and arithmetic expressions, etc. evaluate to the empty sequence
if one of the sides is the empty sequence.

For more complex queries, the FLWOR expression
allows processing each item of a sequence individually.
It consists of an arbitrary number of \mintinline{xquery}{for},
\mintinline{xquery}{let}, \mintinline{xquery}{where},
\mintinline{xquery}{order by}, \mintinline{xquery}{group by},
and \mintinline{xquery}{return} clauses in (almost) arbitrary order.
The fact that FLWOR expressions
(like any other expression) can be nested arbitrarily
allows users to use the same language constructs
for any nesting level of the data.
For example, the following query
extracts the commits from every push event
that were authored by the committer who authored most of them:
\vspace{1ex}
\begin{minted}{xquery}
for $e in $events         (: iterate over input :)
let $top-committer := (
  for $c in $e.commits[]  (: iterate over array :)
  group by $c.author
  stable order by count($c) descending
  return $c.author)[1]
return [$e.commits[][$$.author eq $top-committer]]
\end{minted}
\vspace{1ex}
Appendix~\ref{app:commits-by-top-committers}
shows our attempt to formulate the above query in SQL,
which is about five times longer than that in JSONiq.

Other expressions include literals, function calls,
sequence predicates, sequence concatenation,
range construction, and various ``data-flow-like'' expressions
such as \texttt{if-then-else}, \texttt{switch},
and \texttt{try-catch} expressions.
Of particular interest for JSON data sets
are the expressions for object and array creation,
array access, and merging of objects,
as well as expressions dealing with types such as \texttt{instance-of},
\texttt{castable} and \texttt{cast}, and \texttt{typeswitch}.
Finally, JSONiq comes with a rich function library
including the large function library standardized by W3C.

While the ideal query language heavily depends
on the taste and experience of the programmer,
we believe that the language constructs summarized above
for dealing with absent, nested, and heterogeneous data
make JSONiq very well suited for querying large, messy data sets.
In the experimental evaluation, we also show
that the higher programming comfort of that language
does not need to come with a large performance penalty.

\section{Design and Implementation}

\subsection{Challenges and Overview}

The high-level goal of Rumble is to execute arbitrary JSONiq queries
on data sets of any scale that technology permits.
We thus design Rumble as an application on top of Spark
in order to inherit its big-data capabilities.

On a high level, Rumble follows a traditional database architecture:
Queries are submitted either individually via the command line or in a shell.
When a query is submitted, Rumble parses the query into an AST,
which it then translates into a physical execution plan.
The execution plan consists of \emph{runtime iterators},
which each, roughly speaking, correspond
to a particular JSONiq expression or FLWOR clause.
Finally, the result is either shown on the screen of the user,
saved to a local file,
or, if the root iterator is executed in a Spark job,
written to a file by Spark.

The key challenge in this design
is mapping the nested and heterogeneous data model
as well as the fully recursive structure of JSONiq
onto the data-frame-based data model of Spark
and the (mostly flat) execution primitives defined on it.
In particular, since JSONiq uses
the same language constructs for any nesting level,
we must decide which of the nesting levels
we map to Sparks distributed execution primitives.

The main novelty in the design of Rumble is the runtime iterators
that can switch dynamically between different execution modes,
and which encode the decision of which nesting level is distributed.
In particular, iterators have one of the following execution modes:
(i) \emph{local execution},
which is done using single-threaded Java implementations
(i.e., the iterator is executed independently of Spark),
(ii) \emph{RDD-based execution}, which uses Spark's RDD interface,
and (iii) \emph{DataFrame-based execution}, which uses the DataFrame interface.
Typically, iterators of the outer-most expressions use the last two modes,
i.e., they are implemented with higher-order Spark primitives.
Their child expressions are represented
as a tree of iterators using the local execution mode,
which is passed as argument into these higher-order primitives.

In the remainder of this section,
we first describe the runtime iterators in more detail
and then present how to map JSONiq expressions onto these iterators.

\subsection{Runtime Iterators}

We first go into the details of the three execution modes.
\emph{Local execution} consists of Volcano-style iterators
and is used for pre- or post-processing of the Spark jobs
as well as for nested expressions passed as an argument
to one of Spark's higher-order primitives.
Simple queries may entirely run in this mode.
In all of these cases, local execution
usually deals with small amounts of data at the time.
In addition to the usual
\mintinline{Java}{open()}, \mintinline{Java}{hasNext()},
\mintinline{Java}{next()}, and \mintinline{Java}{close()} functions,
our iterator interface also includes \mintinline{Java}{reset()}
to allow for repeated execution of nested plans.
Apart from the usual semantics,
the \mintinline{Java}{open()} and \mintinline{Java}{reset()} functions
also set the dynamic context of each expression,
which essentially provides access to the in-scope variables.

The other two execution modes are based on Spark.
The \emph{Data\-Frame-based execution} is used
whenever the internal structure (or at least part of it) is known statically.
This is the case for the tuple-based iterators of FLWOR clauses,
where the variables bound in the tuples
can be derived statically from the query
and, hence, represented as columns in a DataFrame.
Some expression iterators, whose output structure can be determined statically,
also support this execution mode.
Using the DataFrame interface in Spark is generally preferable
as it results in faster execution.
The \emph{RDD-based execution} is used whenever no structure is known,
which is typically the case for sequences of items.
Since we represent items as instances of a polymorphic class hierarchy,
which is not supported by DataFrames,
we use RDDs instead.

Each execution mode uses a different interface:
local execution uses the
\mintinline{Java}{open()}/\mintinline{Java}{next()}/\mintinline{Java}{close()}
interface described above,
while the interfaces of the other two modes
mainly consist of a \mintinline{Java}{getRDD()}
and a \mintinline{Java}{getDataFrame()} function, respectively.
This allows chaining Spark primitives as long as possible
and, hence to run large parts of the execution plan in a single Spark job.
Each runtime iterator has a ``highest'' potential execution mode
(where DataFrame-based execution is considered higher than RDD-based execution,
which, in turn, is considered higher than local execution)
and implements all interfaces up to that interface.
Default implementations make it easy to support lower execution modes:
by default, \mintinline{Java}{getRDD()} converts a DataFrame (if it exists)
into an RDD\footnote{
  This is a simple call
  to the \mintinline{Java}{rdd()} method of Spark's DataFrames.}
and the default implementations of the (local)
\mintinline{Java}{open()}/\mintinline{Java}{next()}/\mintinline{Java}{close()} interface
materialize an RDD (if it exists) and return its items one by one.
Iterator classes may override these functions with more optimal implementations.
The execution modes available for a particular iterator
may depend on the execution modes available for its children.
Furthermore, which interface and hence execution mode is finally used
also depends on the consumer, who generally chooses the highest available mode.

For example, consider the following query:
\begin{minted}{xquery}
count( for $n in json-file("numbers.json")
       return $n )
\end{minted}
The iterator holding the string literal \mintinline{xquery}{"numbers.json"}
only \mbox{supports} local execution,
so the iterator of the built-in function \mintinline{xquery}{json-file()}
uses that mode to consume its input.
Independently of the execution mode of its child,
the \mintinline{xquery}{json-file()} iterator always returns an RDD.
This ensures that reading data from files always happens in parallel.
The iterator of the \mintinline{xquery}{for} clause detects that its child,
the \mintinline{xquery}{json-file()} iterator,
can produce an RDD, so it uses that interface to continue parallel processing.
As explained in more detail in Section~\ref{sec:design:flwor} below,
it produces a DataFrame with a single column holding \mintinline{xquery}{$n}. %
The subsequent \mintinline{xquery}{return} clause
consumes that output through the RDD interface.
The \mintinline{xquery}{return} clause, in turn,
only implements the RDD interface (since it returns a sequence of items).
The \mintinline{xquery}{$n} expression %
nested inside the \mintinline{xquery}{return} clause
is applied to every tuple in the input DataFrame
using the local execution mode for all involved iterators.
Next, the iterator of the built-in function \mintinline{xquery}{count()}
detects that its child iterator (the \mintinline{xquery}{return} clause)
can produce an RDD,
so it uses Spark's \mintinline{Python}{count()} function of RDDs
to compute the result.
Since \mintinline{xquery}{count()} always returns a single item,
its iterator only implements the local execution mode,
which is finally used by the execution driver to retrieve the query result.

In contrast, consider the following query:
\begin{minted}{xquery}
count( for $p in 1 to 10
       return $p.name )
\end{minted}
Since the range iterator executing the \mintinline{xquery}{to} expression
only supports local execution,
its parent can also only offer local execution,
which repeats recursively until the root iterator
such that the whole query is executed locally.
In particular and in contrast to the previous query,
the \mintinline{xquery}{count()} iterator now uses local execution,
which consumes its input one by one through \mintinline{Java}{next()} calls
incrementing a counter every time.

To summarize, Rumble can switch seamlessly
between local and Spark-based execution.
No iterator \emph{needs} to know how its input is produced,
whether in parallel or locally,
but \emph{can} exploit this information for higher efficiency.
This allows to nest iterators arbitrarily
while maintaining parallel
and even the distributed DataFrame-based execution wherever possible.

\begin{table}
  \centering
  \begin{tabular}{@{}l@{\hspace{1em}}p{6.2cm}@{}}
    \toprule
    \textbf{Category} & \textbf{Expression/Clause} \\
    \midrule
    local-only        & \mintinline{xquery}{()},
                        \mintinline{xquery}{{$k:$v}},
                        \mintinline{xquery}{[$seq]}, %
                        \mintinline{xquery}{$$}, %
                        \mintinline{xquery}{+},
                        \mintinline{xquery}{-},
                        \mintinline{xquery}{mod},
                        \mintinline{xquery}{div},
                        \mintinline{xquery}{idiv},
                        \mintinline{xquery}{eq},
                        \mintinline{xquery}{ne},
                        \mintinline{xquery}{gt},
                        \mintinline{xquery}{lt},
                        \mintinline{xquery}{ge},
                        \mintinline{xquery}{le},
                        \mintinline{xquery}{and},
                        \mintinline{xquery}{or},
                        \mintinline{xquery}{not},
                        \mintinline{xquery}{$a||$b},
                        \mintinline{xquery}{$f($x)},
                        \mintinline{xquery}{$m to $n},
                        \mintinline{xquery}{try catch},
                        \mintinline{xquery}{cast},
                        \mintinline{xquery}{castable},
                        \mintinline{xquery}{instance of},
                        \mintinline{xquery}{some $x in $y satisfies...} \\
    \midrule
    \makecell[tl]{sequence-\\producing} &
                        \mintinline{xquery}{json-file},
                        \mintinline{xquery}{parquet-file},
                        \mintinline{xquery}{libsvm-file},
                        \mintinline{xquery}{text-file},
                        \mintinline{xquery}{csv-file},
                        \mintinline{xquery}{avro-file},
                        \mintinline{xquery}{root-}
                        \mintinline{xquery}{file},
                        \mintinline{xquery}{structured-json-file},
                        \mintinline{xquery}{parallelize} \\
    \midrule
    \makecell[tl]{sequence-\\transforming} &
                        \mintinline{xquery}{$seq[...]},
                        \mintinline{xquery}{$a[$i]},
                        \mintinline{xquery}{$a[]},
                        \mintinline{xquery}{$a[[]]},
                        \mintinline{xquery}{$o.$s},
                        \mintinline{xquery}{$seq!...},
                        \mintinline{xquery}{annotate},
                        \mintinline{xquery}{treat} \\
    \midrule
    \makecell[tl]{sequence-\\combining} &
                        \mintinline{xquery}{$seq1,$seq2},
                        \mintinline{xquery}{if ($c) then... else...},
                        \mintinline{xquery}{switch ($x) case... default...},
                        \mintinline{xquery}{typeswitch ($x) case... default...} \\
    \midrule
    \makecell[tl]{FLWOR\\expressions} &
                        \mintinline{xquery}{for},
                        \mintinline{xquery}{let},
                        \mintinline{xquery}{where},
                        \mintinline{xquery}{group by},
                        \mintinline{xquery}{order by},
                        \mintinline{xquery}{count} \\
    \bottomrule
  \end{tabular}
  \smallskip
  \caption{Runtime iterator categorization for JSONiq expressions and clauses.}
  \label{tbl:iterator-overview}
\end{table}

The remainder of this section presents
how we map JSONiq expressions to Rumble's runtime iterators.
For the purpose of this presentation, we categorize the iterators
as shown in Table~\ref{tbl:iterator-overview}
based on the implementation techniques we use.
We omit the discussion of the local-only iterators
as well as the local execution mode of the remaining ones
as their implementations are basically translations
of the pseudo-code of their specification to Java.

\subsection{Sequence-transforming Expressions}

A number of expressions in JSONiq transform sequences of items of any length,
i.e., they work on \emph{item*}.
Consider again the example from above:
\begin{minted}{xquery}
$payload.commits[].author
\end{minted}
The object member look-up operator \mintinline{xquery}{.}
and the array unbox operator \mintinline{xquery}{[]}
transform their respective input sequence
by extracting the \mintinline{xquery}{commits} member
and all array elements, respectively.

Rumble implements this type of expressions
using Spark's \mintinline{Java}{map()},
\mintinline{Java}{flatMap()},
and \mintinline{Java}{filter()} transformations.
The function parameter given to these transformations,
which is called on each item of the input RDD,
consists a potentially parameterized closure that is specific to this iterator.
For example, the object look-up iterator uses \mintinline{Java}{flatMap()}
and its closure is parameterized with the member name to look up
(\mintinline{xquery}{"commits"} and \mintinline{xquery}{"author"}
in the example above).
Similarly, the unbox iterator uses \mintinline{Java}{flatMap()}
with an unparameterized closure.
Note that both need to use \mintinline{Java}{flatMap()}
and not \mintinline{Java}{map()}
as they may return fewer or more than one item per input tuple.
The predicate iterator used in \mintinline{xquery}{$seq[...]} %
is implemented with \mintinline{Java}{filter()}
and its closure is parameterized with a nested plan of runtime iterators,
which is called on each item of the input RDD.

Another noteworthy example in this category
is the JSONiq function \mintinline{xquery}{annotate()},
which ``lifts'' an RDD to a DataFrame given a schema definition by the user.
It is implemented using \mintinline{Java}{map()}
and its closure attempts to convert each instance of \mintinline{Java}{Item}
to an object with the members and their types from the given schema,
thus allowing for more efficient execution.

\subsection{Sequence-producing Expressions}

Rumble has several built-in functions
that trigger distributed execution from local input:
the functions reading from files of various formats
as well as \mintinline{xquery}{parallelize()}.
All of them take a local input
(the file name and potentially some parameters or an arbitrary sequence)
and produce a DataFrame or RDD.

\mintinline{xquery}{parallelize()} takes any local sequence
and returns an RDD-based iterator with the same content.
Semantically, it is thus the identity function;
however, it allows manually triggering
distributed execution from local input.
An optional second argument to the function
allows setting the number of Spark partitions explicitly.
The implementation of this iterator
is essentially a wrapper around Spark's function with the same name.

The two functions \mintinline{xquery}{json-file()}
and \mintinline{xquery}{text-file()}
read the content of the files matched by the given pattern
either parsing each line as JSON object
(i.e., reading JSON lines documents)
or returning each line as string, respectively.
They are RDD-based as no internal structure is known.
Both are essentially wrappers around Sparks \mintinline{Java}{textFile()},
but the iterator of \mintinline{xquery}{json-file()}
additionally parses each line
into instances of Rumble's polymorphic \mintinline{Java}{Item} class.

Finally, a number of structured file formats
are exposed through built-in functions that are based on DataFrames:
Avro, CSV, libSVM, JSON (using Spark's schema discovery), Parquet, and ROOT.
All of them have readers for Spark that produce DataFrames,
which expose their ``columns'' or ``attributes.''
Their implementations are again mostly wrappers around existing functionality
and support for more formats can be added easily based on the same principle.

\subsection{Sequence-combining Expressions}

Several operators in JSONiq combine two or more sequences.
The simplest one is the concatenation operator \mintinline{xquery}{,} ,
which we simply translate into \mintinline{java}{RDD.union}.

Most of the remaining ones contain some sort of control flow,
which allows for a simple optimization:
For example, in the conditional expression
\mintinline{xquery}{if ($c) then... else...},
we first eagerly evaluate the expression in the condition
(\mintinline{xquery}{$c} in the example)
using the local interface.
This may trigger a Spark job, namely if some child expression
uses a Spark-based execution mode.
We then return the RDD of the corresponding branch.
This makes the evaluation of the conditional branches \emph{lazy}:
only the one that is taken is ever submitted to Spark.
The same technique also works
for \mintinline{xquery}{switch} and \mintinline{xquery}{typeswitch}.

\subsection{FLWOR Expressions}
\label{sec:design:flwor}

FLWOR expressions are probably the most powerful expressions in JSONiq
and roughly correspond to SQL's \mintinline{SQL}{SELECT} statements.
They consist of sequences of clauses,
where each clause except \mintinline{xquery}{return}
can occur any number of times,
with the restrictions that the first clause must be
either \mintinline{xquery}{let} or \mintinline{xquery}{for}
and the last clause must be \mintinline{xquery}{return}.

The specification describes the semantics of the FLWOR expression
using \emph{streams} of \emph{tuples}.
A tuple is a binding of sequences of items to variable names
and each clause passes a stream of these tuples to the next.
The initial clause takes a stream of a single empty tuple as input.
Finally, as explained in more detail below,
the \mintinline{xquery}{return} clause
converts its tuple stream to a sequence of items.
In Rumble, we represent tuple streams as DataFrames,
whose columns correspond to the variable names in the tuples.

\subsubsection{For Clauses}
\label{section-data-frames-mapping}

The \mintinline{xquery}{for} clause is used for iteration through a sequence.
It returns a tuple in the input stream
for each item in the sequence
where that item is bound to a new variable.
If the \mintinline{xquery}{for} clause
does not use any variable from a previous clause,
it behaves like the \mintinline{SQL}{FROM} clause in SQL,
which, if repeated, produces the Cartesian product.
However, it may also recurse into local sequences
such as in the example seen above:
\begin{minted}{xquery}
for $c in $e.commits[]  (: iterate over array in $e :)
\end{minted}

In the case of the first \mintinline{xquery}{for} clause
(i.e., there were either no preceding clauses at all
or only \mintinline{xquery}{let} clauses),
the resulting tuple stream
simply consists of one tuple per item of the sequence
bound to the variable name introduced in the clause.
In this case, the runtime iterator of the clause
simply forwards the RDD or DataFrame of the sequence,\footnote{
  As mentioned earlier, we only describe
  the RDD-based and DataFrame-based execution modes
  to keep the exposition concise.
  If the input sequence is provided by an iterator
  that only supports local execution,
  the \mintinline{xquery}{for} clause as well as the other clauses
  continue in local execution as well.}
lifting it from RDD to DataFrame if necessary
and renaming the columns as required.

Subsequent \mintinline{xquery}{for} clauses are handled differently:
First, the expression of the \mintinline{xquery}{for} clause is evaluated
for each row in the input DataFrame (i.e., for each tuple in the input stream)
using a Spark SQL user-defined function (UDF).
The UDF is a closure parameterized
with the tree of runtime iterators of the expression,
takes all existing variables as input,
and returns the resulting sequence of items as a \mintinline{Java}{List<Item>}.
In order to obtain one tuple per input tuple
\emph{and} item of those sequences,
we use Spark's \mintinline{Java}{EXPLODE} functionality,
which is roughly equivalent to \mintinline{Java}{flatMap()} on RDDs.

For example, if the current variables
are \mintinline{SQL}{x}, \mintinline{SQL}{y}, and \mintinline{SQL}{z},
and the new variable introduced by the \mintinline{xquery}{for} clause
is \mintinline{SQL}{i},
then the \mintinline{xquery}{for} clause is mapped to the following:
\begin{minted}{SQL}
SELECT x, y, z, EXPLODE(UDF(x, y, z)) AS i
FROM input_stream
\end{minted}
where \mintinline{SQL}{input_stream} refers to the input DataFrame
and \mintinline{SQL}{UDF} is the UDF described above.

\subsubsection{Let Clauses}
The \mintinline{xquery}{let} clause is used
to bind a new variable to a sequence of items.
Thus, the let clause simply extends each incoming tuple
to include the new variable alongside the previously existing ones.
Similarly to the \mintinline{xquery}{for} clause,
we implement the \mintinline{xquery}{let} clause with a UDF
that executes the iterator tree of the bound expression on each input tuple:
\begin{minted}{SQL}
SELECT x, y, z, UDF(x, y, z) AS i FROM input_stream
\end{minted}

The \mintinline{xquery}{let} and \mintinline{xquery}{for} clauses
allow overriding existing variables.
Formally, they introduce a new variable with the same name,
but since this makes the old variable inaccessible,
we can drop the column corresponding to the latter
from the outgoing DataFrame.

\subsubsection{Where Clauses}

The \mintinline{xquery}{where} clause filters the tuples in the input stream,
passing on only those where the given expression
evaluates to \mintinline{xquery}{true}.
Similarly to the clauses discussed above,
we wrap that expression into a UDF and convert the resulting value to Boolean%
\footnote{More precisely, we take the effective Boolean value.}
and use that UDF in a \mintinline{SQL}{WHERE} clause:
\begin{minted}{SQL}
SELECT x, y, z
WHERE UDF(x, y, z)
FROM input_stream
\end{minted}

\vspace{1ex}
\subsubsection{Group-by Clauses}
\label{section-group-by-mapping}

The \mintinline{xquery}{group by} clause
groups the tuples in the input stream by a (possibly compound) key
and concatenates the sequences of items
of all other variables of the tuples in the same group.
This is similar to SQL's \mintinline{SQL}{GROUP BY} clause
but with the major difference that it also works without aggregation.
Furthermore, the order of the groups is defined by the specification
and the items in the grouping variables
may be of different (atomic) types,
each type producing its own group.
For example, consider the following query:
\vspace{1ex}
\begin{minted}{xquery}
for $x in (1, 2, 2, "1", "1", "2", true, null)
group by $y := $x
return {"key": $y, "content": [$x]}
\end{minted}
\vspace{1ex}
The result is six objects with the values
\mintinline{xquery}{1}, \mintinline{xquery}{2},
\mintinline{xquery}{"1"}, \mintinline{xquery}{"2"},
\mintinline{xquery}{true}, and \mintinline{xquery}{null}
in \mintinline{xquery}{"key"}
and an array with the repeated items in \mintinline{xquery}{"content"}.

In short, we map this clause
to a \mintinline{SQL}{GROUP BY} clause in Spark SQL
and use \mintinline{SQL}{COLLECT_LIST}
to combine all items of each group into one array.%
\footnote{Interestingly, similar approaches
          are used for provenance tracking
          \cite{Niu2017,Psallidas2018,Glavic2009}.}
However, this has several challenges.

First, Spark can only group by atomic, statically typed columns
but the grouping variables in JSONiq are polymorphic items.
We solve this problem by shredding the items in the grouping variables
into three new columns.
Since the grouping variables must be atomic types,
there are only a few cases to consider:
any number, Boolean, and \mintinline{xquery}{null}
can be stored in a \mintinline{SQL}{DOUBLE} column
while strings, the only remaining atomic type,
can be stored in a \mintinline{SQL}{VARCHAR} column.
Finally, we add a third column with an enum indicating the original type.
This way, we can use the three new columns of each grouping variable
as grouping variables in the \mintinline{SQL}{GROUP BY} clause.
The three columns are computed by a dedicated UDF
similar to what we do in the other clauses.
We also add an \mintinline{SQL}{ORDER BY} clause on the same attributes
to sort the groups as mandated by the specification.\footnote{
  \mintinline{xquery}{null} may be configured to be before or after all types,
  which we achieve by defining two values in the enum type,
  one before and one after all other types,
  and pick the appropriate one at runtime.}

Second, we need to recover the original items in the grouping variables.
Since they cannot be grouping columns in SQL,
we must use them in an aggregate function.
All values of columns of the original grouping variables in one group
must be the same by definition,
so we can just pick the first one
using the aggregate function \mintinline{SQL}{FIRST}.
Finally, to keep the interface to the subsequent DataFrame simpler,
we also store these as SQL arrays (which always contain one element).

Putting everything together, consider the following example query,
where \mintinline{SQL}{x} and \mintinline{SQL}{y} are the grouping variables,
\mintinline{SQL}{z} is the only non-grouping variable,
and \mintinline{SQL}{UDF} the UDF used for shredding items:
\begin{minted}{SQL}
SELECT ARRAY(FIRST(x)),
       ARRAY(FIRST(y)),
       COLLECT_LIST(z)
GROUP BY UDF(x), UDF(y)
ORDER BY UDF(x), UDF(y)
FROM input_stream
\end{minted}
\vspace{1ex}

As an optimization, Rumble detects if a non-grouping variable
is used in an aggregate function such as \mintinline{xquery}{count()}
by one of the subsequent clauses
and maps it to the corresponding aggregate function in SQL.
Similarly, non-grouping variables that are not used at all are dropped.

\vspace{1ex}
\subsubsection{Order-by Clauses}

The \mintinline{xquery}{order by} clause returns the input tuple stream
sorted by a specific variable or variables.
We use a similar technique as for the sorting
in the \mintinline{xquery}{group by} clause;
however, care must be taken to fully respect the semantics of the language.
The specification mandates
that the sequences in the sorting variables must all be comparable;
otherwise, an error must be thrown.
Sequences are only comparable if
(i) they contain at most one item and
(ii) those items are of the same atomic type or \mintinline{xquery}{null}.
The empty sequence is always comparable
and the user can choose if it is greater or smaller than any other atomic item.

To implement this semantic,
we do a first pass over the input to discover the types
and throw an error if required.
Then we shred the sorting attributes as described above,
omitting either the column for strings or that for numbers
as only one of them is used.

\vspace{1ex}
\subsubsection{Count Clauses}

The \mintinline{xquery}{count} clause introduces a new variable
enumerating each tuple with consecutive numbers starting with 1.
There is currently no functionality in Spark SQL
that enumerates rows of an entire DataFrame with consecutive numbers,
but practitioners have devised an algorithm~\cite{Glotov2018}
based on \texttt{MONOTONICALLY\_\allowbreak{}INCREASING\_ID()}.
This function enumerates the rows \emph{within} each partition
leaving gaps in between them.
The algorithm consists in enumerating
the rows within each partitions using this function,
then computing the partition sizes and their prefix sum,
and finally broadcasting that prefix sum back into the partitions
to correct for the gaps.
This approach is purely based on DataFrames, runs all phases in parallel,
and does not repartition the bulk of the data.

\vspace{1ex}
\subsubsection{Return Clauses}

The \mintinline{xquery}{count} clause ends a FLWOR expression
by converting the tuple stream into the one flat sequence of items
given by an expression evaluated for each tuple.
We evaluate that expression with a UDF as described above,
convert the DataFrame to an RDD,
and unnest the sequences using \mintinline{Java}{flatMap()}.

\subsection{Current Limitations}

With the above techniques, Rumble is able to cover
the majority of JSONiq.
The only two major missing features
are windows in \mintinline{xquery}{for} clauses,
which we plan to integrate soon using similar features as described above,
and updates and scripting, which are on the longer-term agenda.

\pagebreak
\section{Experiments and Results}
\label{section-results}

We now evaluate Rumble on real-world data sets
and compare it to other systems built for large, messy data sets.

\begin{figure*}
  \begin{subfigure}[t]{.255\linewidth}
    \includegraphics[scale=.7]{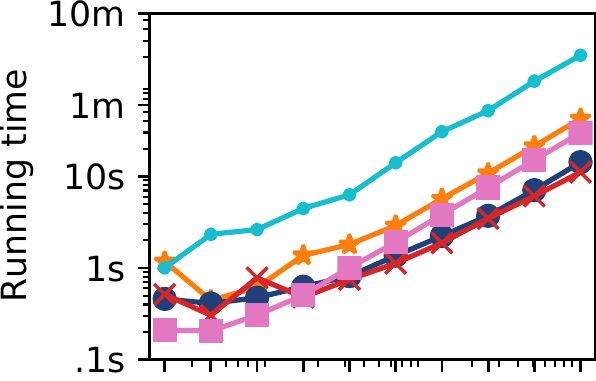}%
    \hfill
    \caption{\textsc{WeatherCount}\hspace*{-3em}}
    \label{fig:cluster:weather-count-star}
  \end{subfigure}
  \begin{subfigure}[t]{.20\linewidth}
    \centering
    \includegraphics[scale=.7]{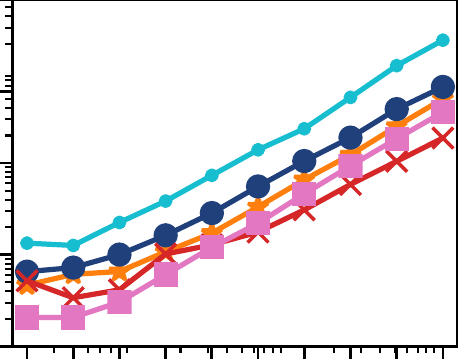}
    \caption{\textsc{WeatherQ0}}
    \label{fig:cluster:weather-q00}
  \end{subfigure}
  \begin{subfigure}[t]{.20\linewidth}
    \centering
    \includegraphics[scale=.7]{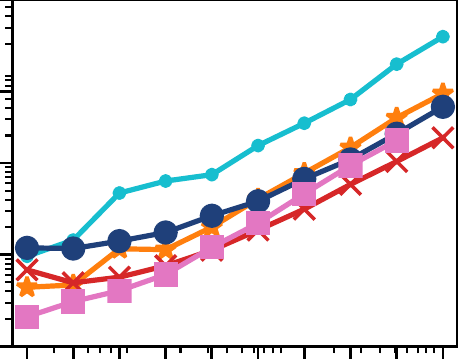}
    \caption{\textsc{WeatherQ1}}
    \label{fig:cluster:weather-q01}
  \end{subfigure}
  \begin{subfigure}[t]{.325\linewidth}
    \hspace{.5em}%
    \includegraphics[scale=.7]{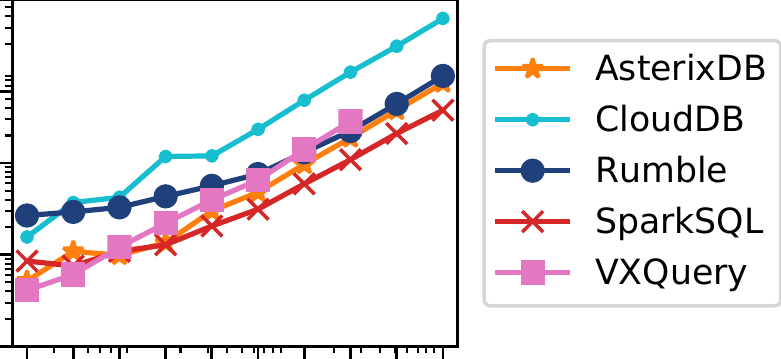}
    \hfill
    \caption{\textsc{WeatherQ2}\hspace*{7em}}
    \label{fig:cluster:weather-q02}
  \end{subfigure}
  \hspace*{\fill}
  \\
  \begin{subfigure}[t]{.255\linewidth}
    \includegraphics[scale=.7]{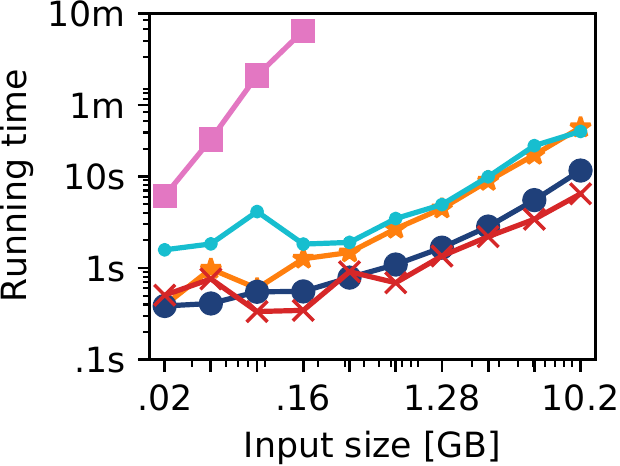}
    \hfill
    \caption{\textsc{GithubCount}\hspace*{-3em}}
    \label{fig:cluster:github-count-star}
  \end{subfigure}
  \begin{subfigure}[t]{.20\linewidth}
    \centering
    \includegraphics[scale=.7]{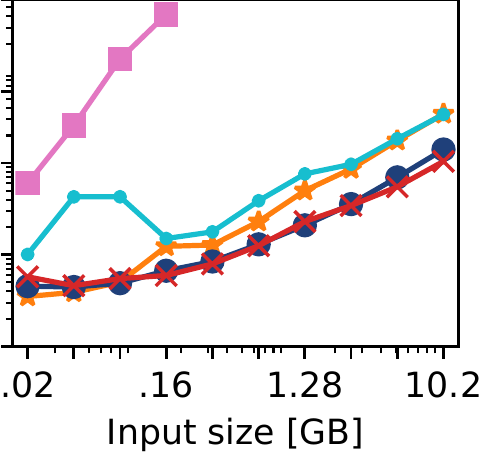}
    \caption{\textsc{GithubFilter}}
    \label{fig:cluster:github-filter}
  \end{subfigure}
  \begin{subfigure}[t]{.20\linewidth}
    \centering
    \includegraphics[scale=.7]{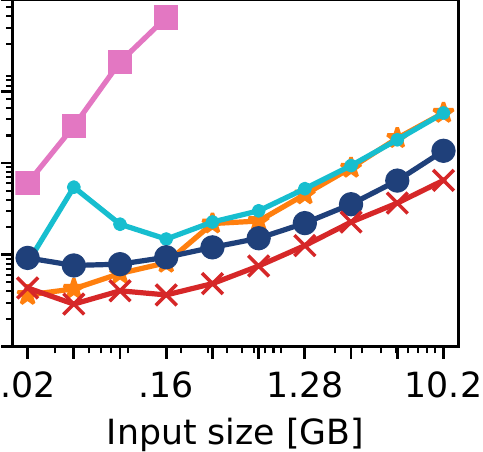}
    \caption{\textsc{GithubGrouping}}
    \label{fig:cluster:github-grouping}
  \end{subfigure}
  \begin{subfigure}[t]{.325\linewidth}
    \hspace{.5em}%
    \includegraphics[scale=.7]{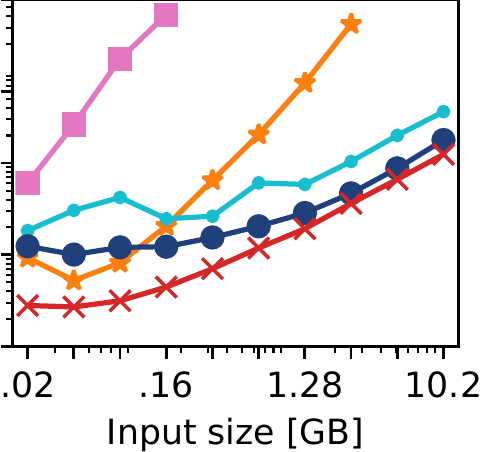}
    \hfill
    \caption{\textsc{GithubSorting}\hspace*{8em}}
    \label{fig:cluster:github-sorting}
  \end{subfigure}
  \hspace*{\fill}\\
  \vspace{-1ex}
  \caption{Performance comparison of distributed JSON processing engines.}
  \label{fig:cluster}
\end{figure*}

\subsection{Experimental Setup}
\label{section-queries}

\noindent\textbf{Platform.}~~%
We conduct all experiments on \texttt{m5} and \texttt{m5d} instances in AWS EC2.
These instances run on machines with Intel Xeon Platinum 8124M CPUs or similar,
have \SI{8}{\gibi\byte} of main memory per CPU core,
and, for the instance sizes we use,
get up to \SI{10}{\mega\bit\per\second} of network bandwidth.
The \texttt{m5d} instances also have a local NVMe SSD.
The \texttt{large} size of each instance type has a one physical CPU core,
the \texttt{xlarge} size has two.
Unless otherwise mentioned, we process the data directly off of S3.

\noindent\textbf{Data sets.}~~%
We use two data sets:
the Github Archive~\cite{GhArchive2020} mentioned before
as well as the \emph{Weather} data set
used by \textcite{pavlopoulou2018} for evaluating VXQuery.
The latter is derived from the
\emph{Daily Global Historical Climatology Network}
(GHCN-Daily) data set~\cite{Menne2012},
which consists of a number of weather-related metrics
from about \SI{115}{\kilo\nothing} weather stations
for each day from 1852 to today.
The data set is published in the domain-specific DLY format;
we converted it to JSON lines
using the scripts provided by \textcite{pavlopoulou2018}.
All records are of the following form:
\begin{minted}{json}
{"data":{"date":"1949-07-01T00:00:00.000","value":0,
   "dataType":"SNWD","station":"GHCND:ACW00011604"}}
\end{minted}
While the latter data set is fully homogeneous and does, hence,
not exercise the strengths of JSONiq,
it allows for comparing our numbers with those from the original authors.
Since not all systems support reading from compressed files,
we store the data uncompressed.

\noindent\textbf{Queries.}~~%
We use the following queries.
For all queries that return more than 20 rows,
we modify the query such that it just counts the number of rows
or specify a limit on the result size.
\\\hspace*{1em}\textsc{(Weather|Github)Count:}~~Count the number of records.
\\\hspace*{1em}\textsc{WeatherQ0:}~~Return the records of measurements
  taken on a December 25 in 2003 or later.
\\\hspace*{1em}\textsc{WeatherQ1:}~~For each date,
  return the number of stations
  that report the \mintinline{json}{"TMIN"} metric.
\\\hspace*{1em}\textsc{WeatherQ2:}~~Compute the average difference
  of the \mintinline{json}{"TMAX"} and \mintinline{json}{"TMIN"} metrics
  reported by the same station on the same day.
\\\hspace*{1em}\textsc{GithubFilter:}~~For each
  \mintinline{json}{"ReleaseEvent"} whose release is a pre-release,
  return the login of the release author.
\\\hspace*{1em}\textsc{GithubGrouping:}~~For each event type,
  count the number of events of that type.
\\\hspace*{1em}\textsc{GithubSorting:}~~Return the actor of each event
  and sort them by their login.

\smallskip

Our experiments are fully automated,
including the generation of the data sets, deployment of the systems in EC2,
executing the queries, and summarizing the results;
the scripts are published in a dedicated source code repository.%
\footnote{\url{https://github.com/RumbleDB/experiments-vldb21}}

\subsection{Comparison of Distributed Engines}

We first compare Rumble with four other query processing engines
for large, messy data sets,
i.e., cluster-based systems that can process JSON data in-situ:
We use Apache AsterixDB, version 0.9.5,
a system for managing and querying heterogeneous data sets;
a commercial cloud-native database system that we refer to as ``CloudDB;''
SparkSQL running on Spark 3;
and VXQuery, rev. \texttt{33b3b79e},
a former Apache project built for analyzing XML and JSON data sets
that is now abandoned.
For Rumble, we use version 1.8.1 on top of Spark 3.
For a fair comparison, we use clusters that cost about \SI{2.5}{\$\per\hour}
in the \texttt{eu-west-1} region:
nine \texttt{m5.xlarge} instances (the default instance type)
on Elastic MapReduce (EMR) for Rumble and SparkSQL,
ten \texttt{m5.xlarge} instances (outside of EMR) for AsterixDB,
and 20 \texttt{m5d.large} instances for VXQuery.
We use a larger number of single-core instances for VXQuery
because we were not able to make it use more than one core per machine.
For CloudDB, we use the \texttt{XSMALL} warehouse size.

\smallskip\noindent
\textbf{Usability.}~~%
We express the queries for AsterixDB, CloudDB, and SparkSQL
in their respective non-standard SQL dialect.
The (limited) support of these dialects for nested data is enough
to express the queries used in this section.
However, SparkSQL does not load the objects in the \texttt{.actor} path
due to a few string values at that path,
so we need to use \mintinline{sql}{FROM_JSON(.)} for inline JSON parsing.
VXQuery uses the JSONiq extension for XQuery,
which should make it easy to deal with messy data sets.
However, we had to try many different reformulations of the queries
to work around bugs in the query engine leading to wrong results
(even for the queries proposed by the original authors),
and were not able to find a correct formulation
for the queries on the Github data set (including \textsc{GithubCount}).
The SQL-based systems require the definition of external tables
before any query can be run.
Except for Spark,
where this takes about twice as long as most of the queries,
this is just a metadata operation that returns immediately.
For VXQuery, the files must be copied manually into the local file system
of the machines in the cluster (which we place on the local SSDs).
Rumble can use the full expressiveness of JSONiq
and query files on cloud storage without prior loading or setup.

\smallskip\noindent
\textbf{Performance.}~~%
Figure~\ref{fig:cluster} shows the running time
of the different systems on the eight queries
for different subsets of the data sets.
We stop all executions after \SI{10}{\minute}.
Most systems are in a similar ballpark for most queries,
in particular for the Weather data set.
They all incur a certain setup overhead for small data sets,
which is expected for distributed query processing engines,
and converge to a stable throughput for larger data sets.
CloudDB is generally among the slower systems;
we assume that this is because
their \texttt{XSMALL} cluster size (which is undisclosed)
is smaller than the clusters of the other systems.
The per-core throughput of VXQuery
is in the order of \SIrange{5}{10}{\mega\byte\per\second},
which corresponds to the numbers in the original paper~\cite{pavlopoulou2018}.
Note that all systems detect the self-join in \textsc{WeatherQ2}
and execute it with a sub-quadratic algorithm.
On the Github data set, some systems show weaknesses:
Curiously, VXQuery has a quadratic running time on all queries
and is, hence, not able to complete queries
on more than \SI{160}{\mega\byte} within the time limit.
(It also crashes for some configurations on the Weather data set.)
Similarly, AsterixDB has a quadratic running time for the sorting query.

Rumble inherits the robustness of Spark
and completes all queries without problems.
As expected, its performance is somewhat lower than that of Spark
due to the polymorphic operators and data representation.
We believe that we can further tighten this gap in the future
by pushing more operations and data representations down to Spark.
Furthermore, the productivity benefits of using JSONiq,
as well as its native support for heterogeneous, nested data sets
make the slightly increased performance cost worthwhile.

\begin{figure}
  \centering
  \includegraphics[scale=.7]{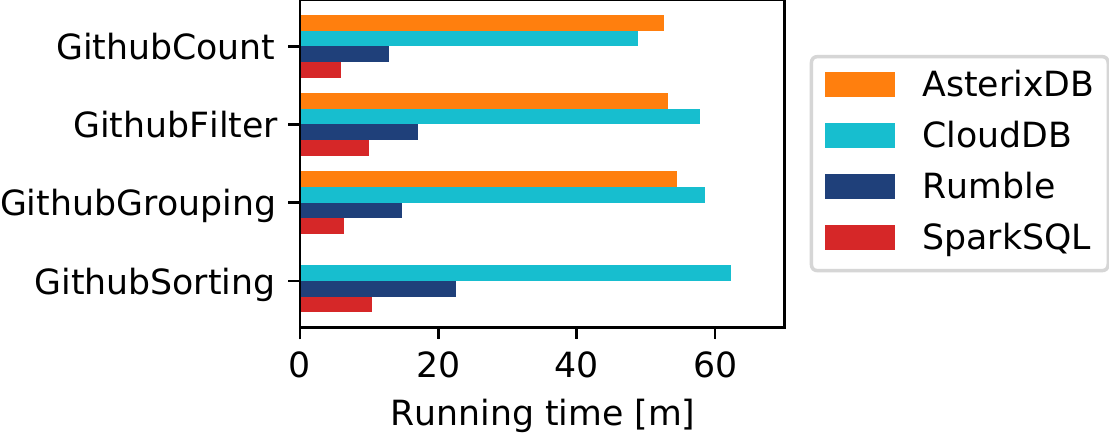}
  \vspace{-1ex}
  \caption{Distributed engines on the full Github data set.}
  \label{fig:github-full}
\end{figure}

\smallskip\noindent
\textbf{Scalability.}~~%
We also run the \textsc{Github*} queries on the full data set,
which is about \SI{7.6}{\tera\byte} large when uncompressed,
and present the results in Figure~\ref{fig:github-full}.
For this experiment, we use clusters that cost about \SI{20}{\$\per\hour},
i.e., that are eight times larger than in the previous experiment.
In order to avoid excessive costs, we run every query only once
and stop the execution after \SI{2}{\hour}.
For the same reason, we compress the input data in this experiment.
CloudDB and AsterixDB are not able to query the data set at this scale
because a tiny number of JSON objects
exceeds the maximum object size
of \SI{16}{\mebi\byte} and \SI{32}{\mega\byte}, respectively.
For CloudDB, we thus report the running time obtained
after removing the problematic objects manually.
This work-around also helps for AsterixDB
but the system then fails with a time-out error.
For reference, we plot extrapolated numbers
from Figure~\ref{fig:cluster} instead.%
\footnote{This results in much more than \SI{2}{\hour}
          for \textsc{GithubSorting},
          which we, hence, omit from the plot.}
We do not include VXQuery here since it is not able
to handle more than \SI{160}{\mega\byte} in the previous experiment.

We observe that the relative performance among the systems remains as before:
Rumble has a moderate performance overhead compared to Spark SQL
but is significantly faster than CloudDB and AsterixDB.
The experiment thus shows
that Rumble can handle the full scale of the data set
in terms of both size and heterogeneity
while providing a high-level language tailored to messy data sets.

\subsection{Comparison of JSONiq Engines}
\label{section-break-even}

\begin{figure*}
 \begin{subfigure}[t]{.255\linewidth}
    \includegraphics[scale=.7]{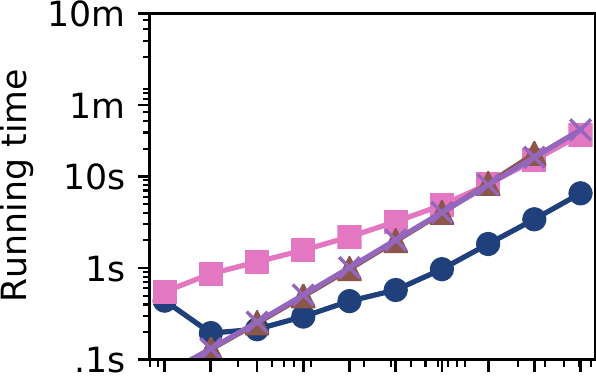}%
    \hfill
    \caption{\textsc{WeatherCount}\hspace*{-3em}}
    \label{fig:singlecore:weather-count-star}
  \end{subfigure}
  \begin{subfigure}[t]{.20\linewidth}
    \centering
    \includegraphics[scale=.7]{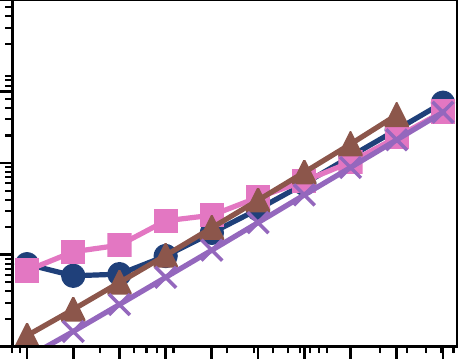}
    \caption{\textsc{WeatherQ0}}
    \label{fig:singlecore:weather-q00}
  \end{subfigure}
  \begin{subfigure}[t]{.20\linewidth}
    \centering
    \includegraphics[scale=.7]{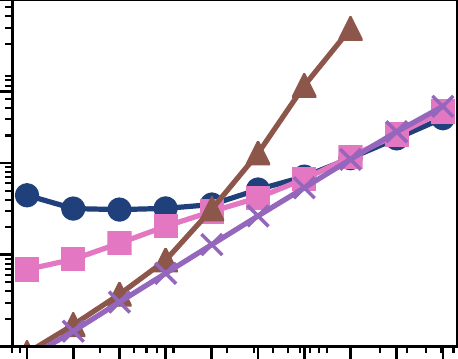}
    \caption{\textsc{WeatherQ1}}
    \label{fig:singlecore:weather-q01}
  \end{subfigure}
  \begin{subfigure}[t]{.325\linewidth}
    \hspace{.5em}%
    \includegraphics[scale=.7]{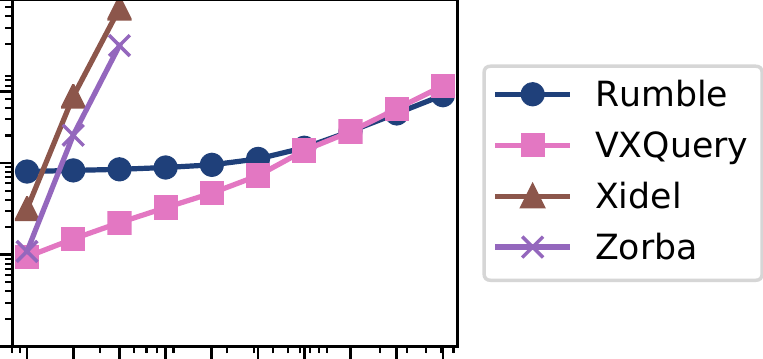}
    \hfill
    \caption{\textsc{WeatherQ2}\hspace*{7em}}
    \label{fig:singlecore:weather-q02}
  \end{subfigure}
  \hspace*{\fill}
  \\
  \begin{subfigure}[t]{.255\linewidth}
    \includegraphics[scale=.7]{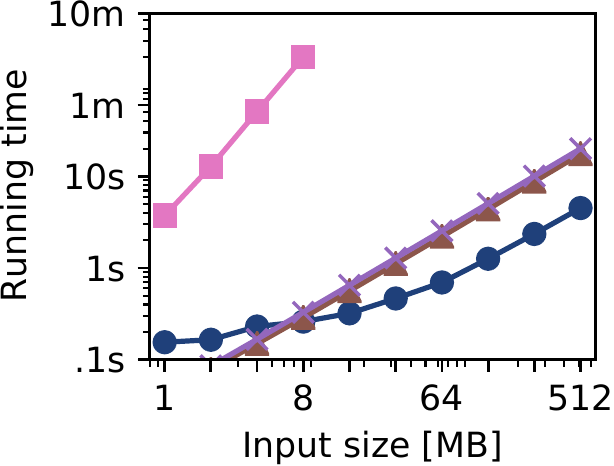}
    \hfill
    \caption{\textsc{GithubCount}\hspace*{-3em}}
    \label{fig:singlecore:github-count-star}
  \end{subfigure}
  \begin{subfigure}[t]{.20\linewidth}
    \centering
    \includegraphics[scale=.7]{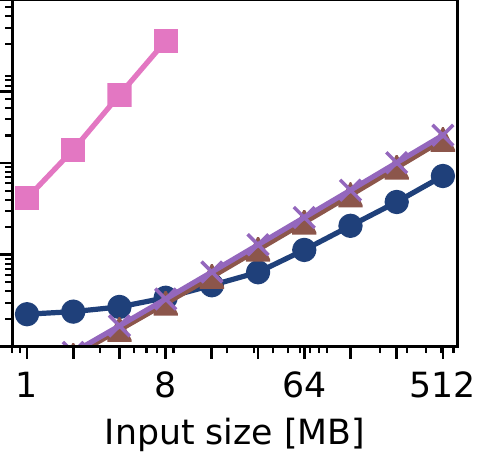}
    \caption{\textsc{GithubFilter}}
    \label{fig:singlecore:github-filter}
  \end{subfigure}
  \begin{subfigure}[t]{.20\linewidth}
    \centering
    \includegraphics[scale=.7]{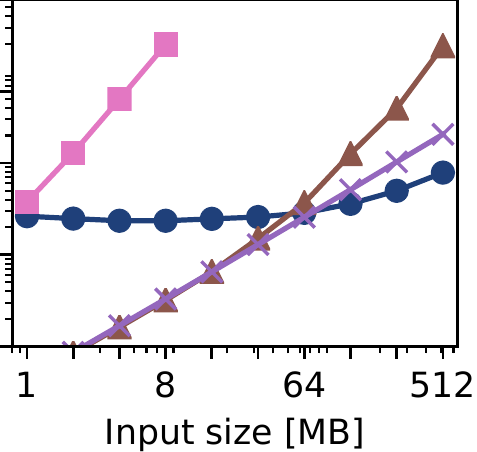}
    \caption{\textsc{GithubGrouping}}
    \label{fig:singlecore:github-grouping}
  \end{subfigure}
  \begin{subfigure}[t]{.325\linewidth}
    \hspace{.5em}%
    \includegraphics[scale=.7]{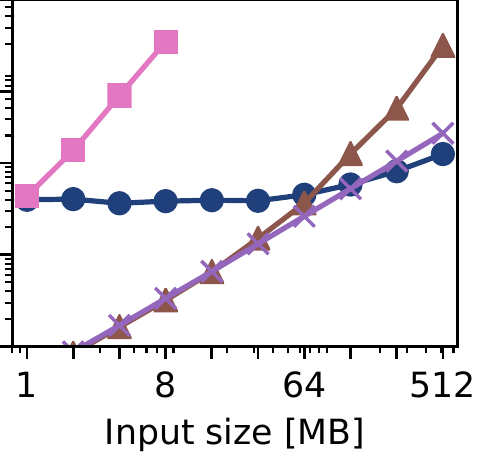}
    \hfill
    \caption{\textsc{GithubSorting}\hspace*{8em}}
    \label{fig:singlecore:github-sorting}
  \end{subfigure}
  \hspace*{\fill}\\
  \vspace{-1ex}
  \caption{Performance comparison of JSONiq engines
           using a single thread.}
  \label{fig:singlecore}
\end{figure*}

In addition to VXQuery, we now compare Rumble with two other JSONiq engines:
Xidel and Zorba.
Xidel is a Pascal implementation; we use version 0.9.8.
Zorba~\cite{zorba} is one of the most complete, stable,
and optimized implementations of JSONiq; we use version 3.0.0.
Both engines are designed for small documents and hence single-threaded.
In order to be able to compare the per-core performance of the engines,
we configure VXQuery and Rumble to use a single thread as well.
Note, however, that is not representative for typical usage
on workstations and laptops,
where the two engines would enjoy a speed-up
roughly proportional to the number of cores.
We run all experiments on \texttt{m5d.large} instances
with the data loaded to the local SSD
and stopped all query executions after \SI{10}{\minute}.

\smallskip\noindent
\textbf{Usability.}~~%
All systems can read files from the local file system,
though with small variations.
Xidel and Zorba use the standardized file module;
VXQuery uses the standardized JSONiq
function \mintinline{xquery}{fn:collection},
and Rumble uses the Rumble-specific function \mintinline{xquery}{json-files}.
Except for those of VXQuery, the remainder of the query implementations
are character-by-character identical between the different systems.
VXQuery has the same limitations in terms of correctness as above;
the other systems behave as expected.

\smallskip\noindent
\textbf{Performance.}~~%
Figure \ref{fig:singlecore} shows the results.
As expected, Xidel and Zorba are considerably faster
than the other two systems on small data sets,
which they are designed and optimized for.
However, they struggle with larger data sets and complex queries:
Xidel cannot run any query on the Weather data set of \SI{512}{\mega\byte}
as it runs out of main memory.
Note that the data should fit comfortably
into the \SI{8}{\gibi\byte} of main memory.
It also has a super-linear running time
for the queries using sorting, grouping, or join.
Zorba can handle more queries,
but, like Xidel, does not seem to have implemented an equi-join
as they have both quadratic running time for \textsc{WeatherQ2}.
VXQuery handles the queries on the Weather data set well;
however, it has the same quadratic running time
on the Github data set as before
and is hence not able to process more than \SI{8}{\mega\byte}
before the timeout.

Rumble can handle all queries well and,
after some start-up overhead for small data sets,
runs as fast as or considerably faster than the other systems.
This confirms that Rumble competes with the per-core performance
of state-of-the-art JSONiq engines,
while at the same time being able to handle
more complex queries on larger data sets.

\section{Related Work}
\label{section-related-work}

\textbf{Query languages for JSON.}
Several languages have been proposed
for querying collections of JSON documents,
a substantial number of them by vendors of document stores.
For example, AsterixDB~\cite{Alsubaiee2014,Alkowaileet2016}
supports the AQL language,
which is maybe the most similar to JSONiq
in the JSON querying landscape.
Other proposals include Couchbase's N1QL~\cite{couchbase,n1ql},
UNQL~\cite{UNQL}, Arango's AQL~\cite{ArangoDB} (homonymous to AsterixDB's),
SQL++~\cite{Ong2014}, and JAQL~\cite{Beyer2011}.
Other languages were proposed for embedding in host language
including PythonQL~\cite{pythonql} and LINQ~\cite{LINQ}.
Finally, MRQL was an alternate proposal
running at scale on MapReduce, Spark, Flink, and Hama,
but was discontinued in 2017.

Most of these and other proposed JSON query languages
address nestedness, missing values, and \mintinline{JSON}{null},
but have only limited support for mixed types in the same path.
As the introductory example motivates,
this frequently happens in real-world use cases
and thus severely limits the applicability of these languages.
We refer to the survey of \textcite{Ong2014}
for an in-depth comparison of these languages.
To the best of our knowledge,
JSONiq is the only language in the survey and among those we mention
that has several independent implementations.

\textbf{Document stores.}
Document stores are related in that they provide native support
for documents in JSON and similar formats~\cite{mongo,esBook,couch}.
Many of them are now mature and popular commercial products.
However, document stores usually target a different use case,
in which retrieving and modifying parts of individual documents
are the most important operations
rather than the analysis of large read-only collections.

\textbf{In-situ data analysis.}
The paradigm employed by Rumble of query data \emph{in-situ}
has received a lot of attention in the past years.
It considerably reduces the time that a scientist needs
in order to start querying freshly received data sets.
Notable systems include NoDB~\cite{NoDB}, VXQuery~\cite{vxQuery},
and Amazon Athena~\cite{Athena}.

\section{Conclusion}

We built Rumble, a stable and efficient JSONiq engine on top of Spark
that provides data independence for heterogeneous, nested JSON data sets
with no pre-loading time.
Our work demonstrates that data independence for JSON processing
is achievable with reasonable performance on top of large clusters.
The decoupling between a logical layer
with a functional, declarative language, on the one hand,
and an arbitrary physical layer
with a low-level query plan language, on the other hand,
enables boosting data analysis productivity
while riding on the coat-tails of the latest breakthroughs
in terms of performance.

Rumble is publicly available as open source since 2018
and has since then attracted users from all over the world.
We use Rumble in the Big Data course at ETH Zürich
and are in contact with a number of institutions that do the same.
Our experience and the feedback of other lecturers
show that, using Rumble, students only need a very short learning period
to do even complex data analyzes on large messy, data sets,
even if their major is not computer science.
In contrast, the corresponding techniques of other tools
(such as \mintinline{SQL}{ARRAY}/\mintinline{SQL}{UNNEST})
either pose more problems
or are judged as ``too complicated'' by the lecturers
and left out completely from their courses.

\printbibliography

\appendix
\section{Commits by Top Committers}
\label{app:commits-by-top-committers}

As argued in the introduction,
most SQL dialects do not allow sub-queries for array construction.
If a dialect does not support that construct,
queries need to repeatedly unnest, group, and/or self-join,
and group with \mintinline{sql}{ARRAY_AGG}
in order to deal with the different nesting levels.
To illustrate how cumbersome queries can become with that approach,
we use it for the sample JSONiq query
given in Section~\ref{sec:background:jsoniq},
which finds the commits of each push event
that were authored by the committer who authored most commits of that push.

\begin{figure}
\begin{minted}{sql}
WITH Commits AS (
  SELECT
    FORMAT('%s-%i', created_at, ROW_NUMBER()
           OVER(PARTITION BY created_at))
      AS event_id,
    payload.shas
  FROM `bigquery-public-data.samples.github_nested`
  WHERE ARRAY_LENGTH(payload.shas) > 0),
CommitterFrequency AS (
  SELECT
    Commits.event_id AS event_id,
    actor_email,
    COUNT(*) commit_count
  FROM Commits, UNNEST(shas)
  GROUP BY event_id, actor_email),
MaxCommitterFrequency AS (
  SELECT
    event_id,
    MAX(commit_count) AS commit_count
  FROM CommitterFrequency
  GROUP BY event_id),
TopCommitters AS (
  SELECT
    c.event_id,
    ANY_VALUE(c.actor_email) AS actor_email
  FROM
    CommitterFrequency c,
    MaxCommitterFrequency m
  WHERE c.event_id = m.event_id AND
    c.commit_count = m.commit_count
  GROUP BY c.event_id),
TopCommitterCommits AS (
  SELECT c.event_id, commits
  FROM
    Commits c,
    UNNEST(shas) AS commits,
    TopCommitters AS tc
  WHERE c.event_id = tc.event_id AND
    commits.actor_email = tc.actor_email
)
SELECT ARRAY_AGG(commits) shas
FROM TopCommitterCommits
GROUP BY event_id
\end{minted}
\caption{``Select commits by top committers'' in SQL.}
\label{fig:commits-by-top-committers-sql}
\end{figure}

\addtolength{\textheight}{-2cm}

A possible implementation (which is fully functional in Google BigQuery)
is given in Figure~\ref{fig:commits-by-top-committers-sql}.
The query first generates a unique ID for each event using \texttt{ROW\_NUMBER}
in order to establish an even identity for subsequent groupings and joins.
To find the frequency of each committer in each \texttt{shas} array,
we unnest commits and group by \mintinline{sql}{event_id, actor_email}
using \mintinline{sql}{COUNT(*)}.
The next three steps are the typical work-around
for SQL's inability to compute \emph{argmin}:
(1) We first group by \mintinline{sql}{event_id}
to find the commit frequency of the top committer per event;
(2) then join the previous two tables
on \mintinline{sql}{event_id, commit_count}
to find the (possibly numerous) committers that match that frequency per event;
(3) and finally group by \mintinline{sql}{event_id} again
to break ties among the equally frequent committers.
To produce the final result,
we unnest all commits again, join them with the top committers per event,
and, to pack the commits of each event back into an array,
group by \mintinline{sql}{event_id} using \mintinline{sql}{array_agg}.

The SQL implementation is 32 lines long and uses six sub-queries
whereas the JSONiq implementation is seven lines long
and uses two FLWOR expressions.
While those are only approximate metrics for capturing the query complexity,
it seems clear that the JSONiq implementation
is considerably more concise and easier to read.

\end{document}